\documentclass[cits]{PoS}

\usepackage{epsfig}
\usepackage{amssymb}
\usepackage{url}
\newcommand{\eqref}[1]{eq.~(\ref{#1})\xspace}

\newcommand{\secref}[1]{Section~\ref{#1}}

\newcommand{\tr}{\ensuremath{\mbox{tr}}}

\newcommand{\Ng}{\ensuremath{N_g\;}}

\newcommand{\qqbar}{\ensuremath{q\overline{q}\;}}
\newcommand{\Nc}{\ensuremath{N_c}}
\usepackage[vcentermath]{youngtab}

\title{Tools for calculations in color space}
\ShortTitle{Tools for calculations in color space}

\author{\speaker{Malin SJODAHL}\\
  Dept. of Astronomy and Theoretical Physics, Lund University, 
  S\"olvegatan 14A, 223\,62~Lund, Sweden, E-mail: \email{malin.sjodahl@thep.lu.se}\\
}

\author{Stefan KEPPELER\\
  Mathematisches Institut, 
  Universit\"at T\"ubingen, 
  Auf der Morgenstelle 10, 
  72076 T\"ubingen, 
  Germany,
  E-mail: \email{stefan.keppeler@uni-tuebingen.de}
}

\abstract{Both the higher energy and the initial state colored partons 
contribute to making exact calculations in QCD color space more important 
at the LHC than at its predecessors. 
This is applicable whether the method of assessing QCD is fixed order 
calculation, resummation, or parton showers. 
In this talk we discuss tools for tackling the problem of performing exact color summed 
calculations. We start with theoretical tools in the form of the 
(standard) trace bases and the orthogonal multiplet bases 
(for which a general method of construction was recently presented). 
Following this, we focus on two new packages for performing 
color structure calculations: one easy to use Mathematica package, 
ColorMath, and one C++ package, ColorFull, 
which is suitable for more demanding calculations, 
and for interfacing with event generators.\\
\begin{picture}(0,0) \put(345,500){\large LU TP 13-25}\end{picture}}

\FullConference{XXI International Workshop on Deep-Inelastic Scattering and Related Subjects\\
		 22-26 April, 2013\\
		 Marseilles, France}

\begin{document}

\section{Introduction}

The high energy available at the LHC and the initial state colored partons 
both contribute to increase the number of colored partons
involved in collisions, and indeed, QCD is a major background for 
most interesting processes. An accurate treatment of QCD and the 
color space associated with it is therefore of increased importance.
This is applicable independent of the method of accessing QCD,
fixed order calculations, resummation, or 
-- as lately for the speaker -- parton showers \cite{Platzer:2012np}.

We here discuss several tools for dealing with exact 
calculations in QCD color space.
After a few words of introduction in \secref{sec:ColorSpace},
we briefly review the standard ``trace bases'' method
in \secref{sec:TraceBases}.

In \secref{sec:MultipletBases}, we present 
recent results on how to instead deal with color space
using orthogonal group theory based multiplet bases
\cite{Keppeler:2012ih}.
Finally, in \secref{sec:ComputerTools}, we 
introduce two computer algebra codes for dealing with 
calculations in color space, the Mathematica package  
ColorMath \cite{Sjodahl:2012nk}, which is designed do 
deal with color summed calculations of moderate complexity
in a user friendly way, and the C++ package
ColorFull \cite{Sjodahl:ColorFull} which aims at 
dealing with trace bases in an event generator context.

\section{Color space}
\label{sec:ColorSpace}

Due to confinement we never observe the color of individual partons in
QCD. This distinguishes color from other internal degrees of freedom
like spin, insofar as for the latter we actually care about in which
state a physical object is, rather than just under what representation
it transforms.

This property also opens up for the possibility to study
summed (averaged) quantities {\it only}. It is not hard to 
argue that for given external partons, 
the color summed (averaged) space is a finite dimensional vector space
equipped with a scalar product 
    \begin{eqnarray}
      \left<A,B\right> = \sum_{a,b,c,...} (A_{a,b,c,...})^* B_{a,b,c,...}\;,
    \end{eqnarray}
where the sum runs over all quarks, anti-quarks and gluons.
For example, if
    \begin{eqnarray}
      A=\sum_g(t^{g})^a\,_{b}(t^{g})^c\,_{d}=\sum_g\parbox{2.5 cm}{\epsfig{file=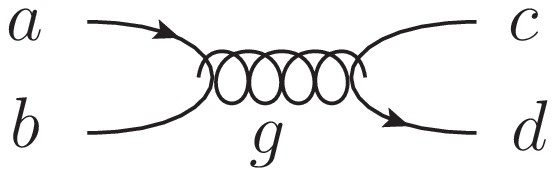,width=2.5cm}}, \nonumber
    \end{eqnarray}
then 
\begin{eqnarray}
  \left<A|A\right>=\sum_{a,b,c,d,g,h} (t^{g})^b\,_{a}(t^{g})^d\,_{c} (t^{h})^a\,_{b}(t^{h})^c\,_{d}\;.
\end{eqnarray}

One way of dealing with the color space is to square the amplitudes 
one by one as they are encountered. Alternatively -- and this is likely
the preferred method for processes with many colored partons --
one may want to use a basis or a spanning set.

\section{Trace bases}
\label{sec:TraceBases}

A standard way of dealing exactly with QCD color space is
to note that every four-gluon vertex can be rewritten in terms of
three-gluon vertices. The three gluon vertices in turn
can be replaced using $i f_{abc}=(1/T_R)[\tr(t^at^bt^c-t^bt^at^c)]$,
where $T_R$ is defined by $\tr[t^at^b]= T_R\delta^{ab}$,
and all internal gluon propagators can be removed using the Fierz or completeness relation,
\begin{eqnarray}
     \parbox{7 cm}{\epsfig{file=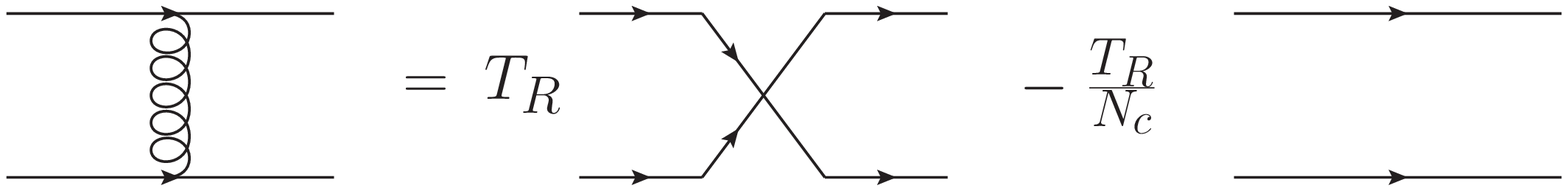, width=7cm}}\;.
\end{eqnarray}

This can be applied to any QCD amplitude, tree level or beyond,
and the result is in general a linear combination of products of
traces over gluon indices and traces that have been cut open, 
i.e. color structures of the form
\begin{eqnarray}
  \parbox{7 cm}{\epsfig{file=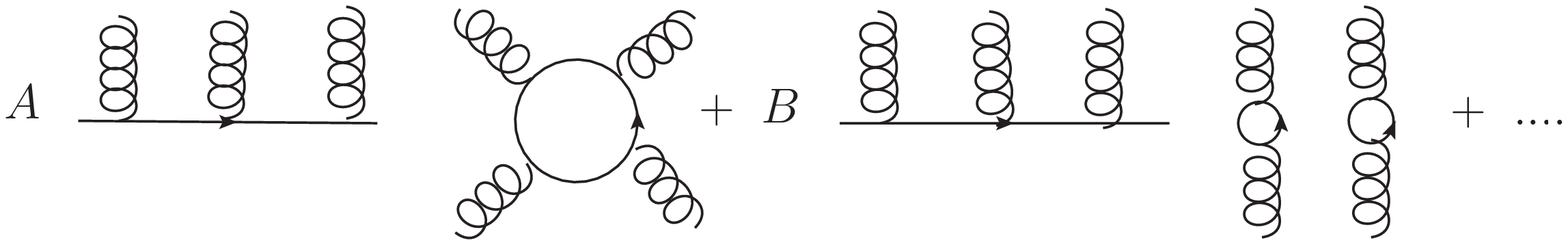,width=7 cm}}\;.
\end{eqnarray}
For obvious reasons this type of basis is here referred to as 
a trace basis.

These bases
\cite{Paton:1969je,Dittner:1972hm,Cvi76,Cvitanovic:1980bu,Mangano:1987xk,Mangano:1988kk,Nagy:2007ty,Sjodahl:2009wx}
have several advantages. It is easy to see  
that a basis vector of this type, results in at most two new
basis vectors (in a larger vector space) once a gluon is emitted.
Furthermore, starting with any basis vector and exchanging a gluon 
between two partons results in a linear combination of at most 
four basis vectors \cite{Sjodahl:2009wx}.
On top of this, powerful recursion relations exist for the
amplitudes multiplying the various color structures.

Trace bases, however, also come with significant drawbacks.
Most importantly, they are not orthogonal, and for more than $\Nc\;$gluons 
plus $\qqbar$-pairs the ``bases'' are also overcomplete. Furthermore,
as the number of spanning vectors in these bases 
grows roughly as a factorial in $\Ng +N_{\qqbar}$ 
this rapidly becomes an issue \cite{Keppeler:2012ih}. 

\section{Multiplet bases}
\label{sec:MultipletBases}

It is therefore desirable to use minimal orthogonal bases.
As QCD is based on SU(3), one way to construct orthogonal bases
is to use bases corresponding to irreducible representations
in color space. Basis vectors where at least one subset of partons
transforms under a different representation will then 
automatically be orthogonal.
One way to enforce this is to sub-group the partons 
in order to make sure that parton 1  and 2 are in a manifest 
multiplet $M^{12}$, at the same time as partons 1,2, and 
3 are in a manifest multiplet $M^{123}$ etc 
\cite{Sjodahl:2008fz,Keppeler:2012ih}.

As the decomposition of color space into irreducible 
representations can be enumerated using Young tableau
multiplication, the expectations on multiplet type 
bases are clear: In general 
(to arbitrary order in perturbation theory),
we expect to encounter any state where the incoming
partons are in multiplet $M$, and the outgoing partons in the 
same multiplet $M$. 
For example, for $qq \to qq$ we have the Young tableau decomposition

{\tiny
  \begin{eqnarray}
    \begin{array}{ccccccccccccccccc}
      \yng(1) &\otimes &\yng(1)& = &\yng(2)& \oplus & \yng(1,1)& \\
\\
      3      &        &  3    &   &  6    &        & {}^{}\overline{3}& \\
    \end{array} \;,\nonumber
  \end{eqnarray}}
and the corresponding orthogonal basis vectors are
\begin{eqnarray}
  & &\parbox{8 cm}{\epsfig{file=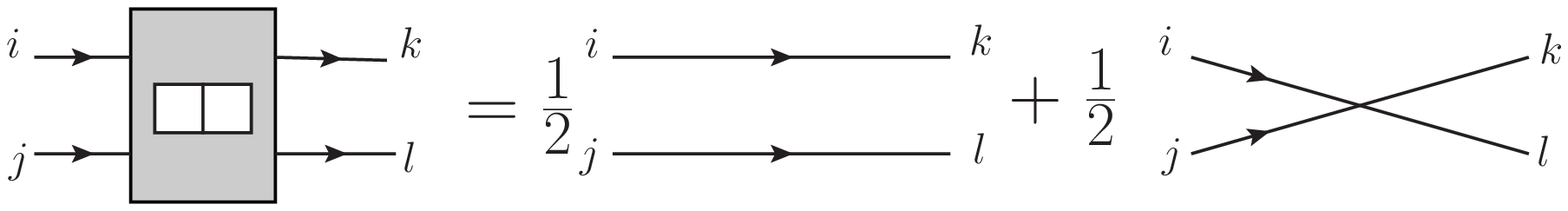,width=8 cm}} 
  = \frac{1}{2}\left( \delta^i{}_k\delta^j{}_l + \delta^i{}_l\delta^j{}_l\right)\\
  & &\parbox{8 cm}{\epsfig{file=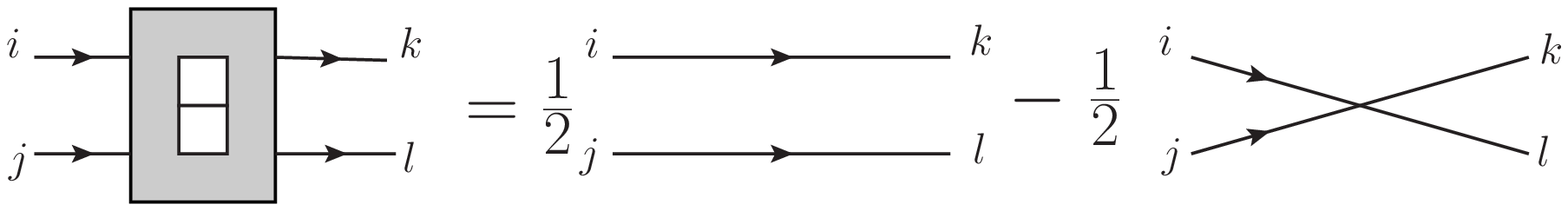,width=8 cm}}
  = \frac{1}{2}\left( \delta^i{}_k\delta^j{}_l - \delta^i{}_l\delta^j{}_l\right)\;.\nonumber
\end{eqnarray}
For processes with only quarks and anti-quarks 
(an incoming anti-quark can always be traded for an outgoing quark etc.,
so we may always treat the color space as if we had $N_q$
incoming quarks and $N_q$ outgoing quarks)
orthogonal bases can be constructed similarly by using 
Hermitian versions of Young 
projection operators \cite{Keppeler:2012ih,Cvitanovic:1980bu,ProjectorPaper}.

For processes with gluons, the translation from Young tableaux to 
basis vectors is far from obvious as the Young tableaux operate
with quark units, rather than gluon units. 
We can enumerate basis vectors using Young tableau multiplication,
the problem lays in the construction of the corresponding basis vectors.

Let us start with considering processes with gluons only.
In the case of $gg \to gg$ the problem of constructing orthogonal bases 
corresponding to the multiplets in 
$8\otimes 8= 1 \oplus 8 \oplus 8 \oplus 10 \oplus \overline{10} +\oplus 27 \oplus 0$
\footnote{For $\Nc>3$ there is an additional multiplet which vanishes for SU($3$).}
has been solved a long time ago 
\cite{MacFarlane:1968vc, Butera:1979na,Cvi84,Cvi08,Dokshitzer:2005ig}. 
Several solutions involve splitting the gluons into
$\qqbar$-pairs and using that for each set of symmetries among the quarks and 
anti-quarks there is precisely one ``new'' multiplet, i.e.
precisely one multiplet that could not occur for fewer gluons. 

For example, the decuplet corresponds to symmetrizing the quarks and anti-symmetrizing the
anti-quarks and can be obtained from the color structure
\begin{eqnarray}
  \mathbf{T}^{10} 
  \sim  \parbox{5cm}{\epsfig{file=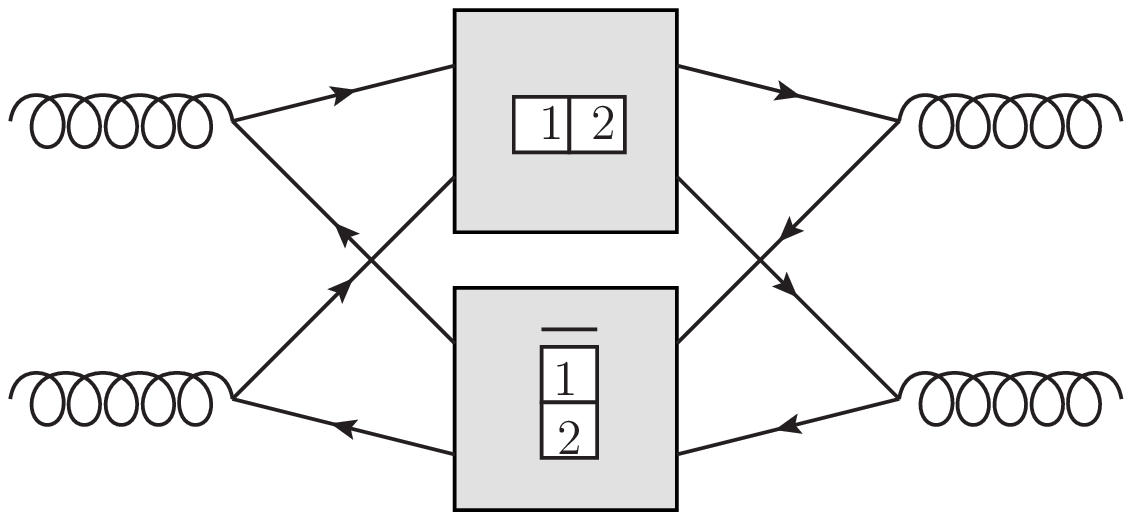,width=5cm}}\;.
\end{eqnarray}
Similarly the anti-decuplet corresponds to
{\tiny $\young(1,2)\otimes \overline{\young(12)}$},
the 27-plet corresponds to {\tiny $\young(12)\otimes \overline{\young(12)}$}
and the 0-plet to {\tiny $\young(1,2)\otimes \overline{\young(1,2)}$}.
By projecting out parts corresponding to ``old'' multiplets, 
i.e., multiplets that can appear also for fewer gluons, 
projection operators and basis vectors can be constructed.
For more than two gluons, the above picture is complicated 
in several ways:

\begin{enumerate}
\item[{\it (i)}] For each (anti-)quark Young diagram (Young tableau shape) 
there are many Young tableaux.

\item[{\it (ii)}] For each new overall multiplet, for example a 
$35$-plet, there are in general several ways of obtaining it, for example 
there is one $35$-plet in $10\otimes 8$ and one in $27 \otimes 8$. 
\end{enumerate}
On top of this, when constructing the basis vectors there are
issues with multiple occurrences of the same multiplet as well as with 
the construction of all vectors corresponding to ``old'' multiplets.

That this method for constructing basis tensors can be fruitful
for more than two gluons therefore appears far from obvious. 
However, to make a long story short, we have shown that 
it is \cite{Keppeler:2012ih}. The proof largely depends on one key observation,
namely that starting in a given multiplet, corresponding to some 
$\qqbar$-symmetries (such as 10, 
from {\tiny$\young(12)\otimes \overline{\young(1,2)}$})
it turns out that for each way of attaching a quark box to 
the quark Young-tableau ({\tiny$\young(12)$}) 
and an anti-quark box to  
the anti-quark Young tableau
({\tiny$\overline{\young(1,2)}$}),
there is at most one new multiplet. For example, the projector
$\mathbf{P}^{10,35}$ can be seen as coming from

\begin{eqnarray}      
 \mathbf{T}^{10,35} \sim \parbox{7cm}{\epsfig{file=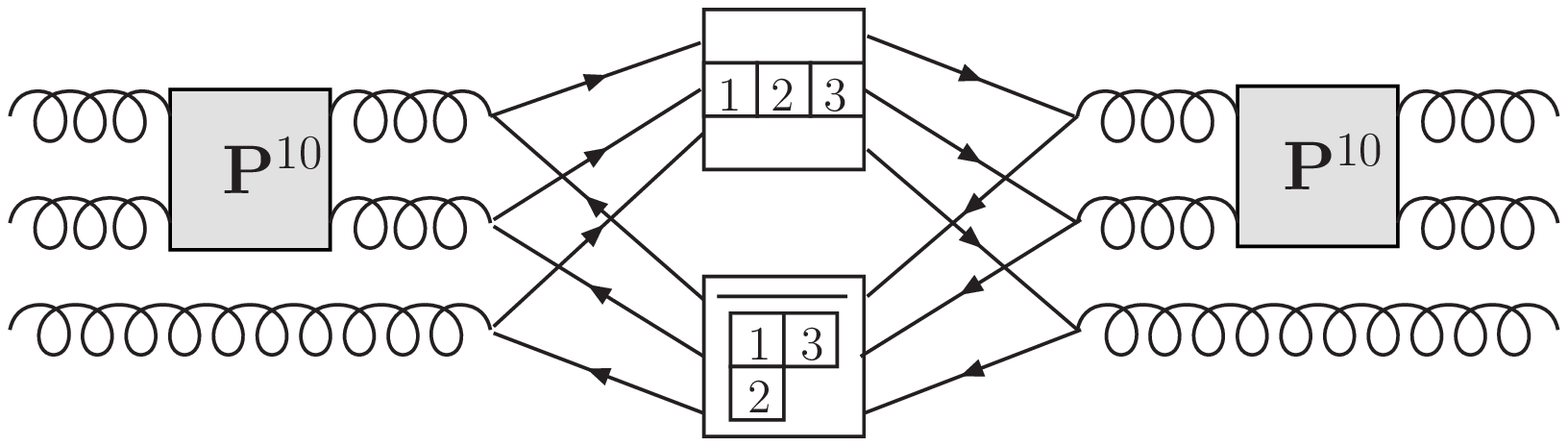, width=7 cm}}  \;,
\end{eqnarray}
after having projected out "old" multiplets.
In fact, for sufficiently large $\Nc$, there is precisely one new multiplet
for each set of $\qqbar$-symmetries. What appears as a problem in 
{\it (i)} is thus in fact the resolution to the problem in {\it (ii)}!

In this way we can construct all projection operators for an arbitrary
number of gluons, and from these we can construct orthogonal minimal 
bases for any number of gluons. 
For the three gluon case, we have explicitly 
constructed all 51 projectors and 265 bases vectors (for general $\Nc$).
The generalization to processes involving both quarks and gluons is 
straightforward, as each $\qqbar$-pair either is in an octet, in which case it can be
treated as a gluon or in a singlet.

\section{Computer tools}
\label{sec:ComputerTools}

In order to facilitate automatic color summed calculations the speaker 
has developed and cross checked two independent computer algebra packages.

\subsection{ColorFull}
For the purpose of treating a general QCD color structure in the trace 
basis, a C++ color algebra code, ColorFull \cite{Sjodahl:ColorFull}, which
creates trace bases for any number and kind of partons and to 
any order in $\alpha_s$ has been written. 
ColorFull also describes the effect of gluon emission and exchange,
squares color amplitudes and is planned to be published separately 
later this year. 

\subsection{ColorMath}

ColorMath \cite{Sjodahl:2012nk} is a user friendly Mathematica package 
for calculations in color space of moderate 
complexity.

In its simplest form, the idea of ColorMath is that one should just write down 
the color structure, much like on paper, and then run {\bf CSimplify} to 
contract color indices, for example

\begin{eqnarray}
\parbox{8 cm}{ \epsfig{file=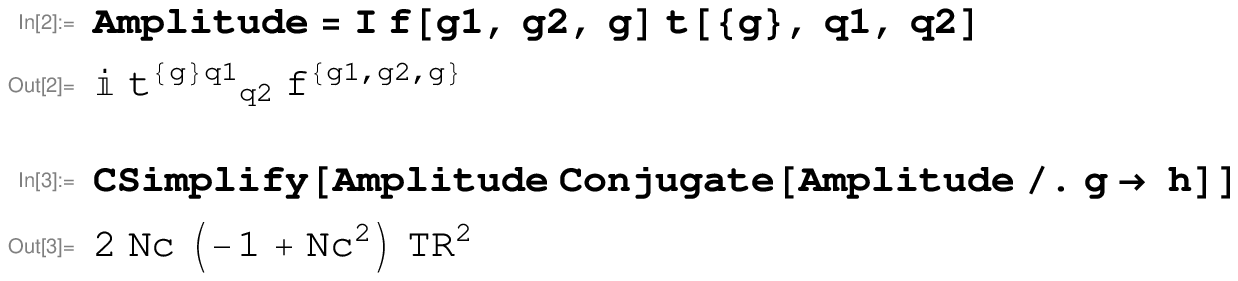, width=8 cm} }\;\;.
\end{eqnarray}

\section{Conclusions}

One way of dealing with exact calculations in color space is to use 
trace bases. This method has advantages when it comes 
to simplicity, recursion relations and the effect of gluon exchange and gluon emission.
It is also the basis for the C++ code ColorFull \cite{Sjodahl:ColorFull},
which is intended for enabling advanced color calculations
in an event generators context.
This type of basis is, however, overcomplete and not orthogonal, which becomes
an issue for many partons due to the rapid growth of the number of spanning vectors.

It is therefore desirable to construct minimal orthogonal bases,
and we have recently outlined a general recipe for group theory based 
minimal orthogonal multiplet bases for any QCD process \cite{Keppeler:2012ih}. 
This can potentially very significantly speed up exact calculations 
in the color space of SU($\Nc$).

We have also presented a Mathematica package ColorMath \cite{Sjodahl:2012nk}
for performing color summed calculations in SU($\Nc$).

\bibliographystyle{JHEP} 
 
\bibliography{Refs}  



\end{document}